\documentstyle[12pt,epsf]{article}

\begin{document}
\begin{titlepage}
\begin{centering}
\vspace{2.0cm}
{\LARGE{\bf Viscous fingering patterns in ferrofluids}}\\
\bigskip\bigskip
Michael Widom and Jos\'e A. Miranda\\
{\em Department of Physics, Carnegie Mellon University, Pittsburgh, PA 15213}\\

\end{centering}

\begin{abstract}
Viscous fingering occurs in the flow of two immiscible, viscous fluids
between the plates of a Hele-Shaw cell. Due to pressure gradients or
gravity, the initially planar interface separating the two fluids
undergoes a Saffman-Taylor instability and develops finger-like
structures. When one of the fluids is a ferrofluid and a
perpendicular magnetic field is applied, the labyrinthine instability
supplements the usual viscous fingering instability, resulting in
visually striking, complex patterns. We consider this problem in a
rectangular flow geometry using a perturbative mode-coupling analysis.
We deduce two general results: viscosity contrast between the fluids
drives interface asymmetry, with no contribution from magnetic forces;
magnetic repulsion within the ferrofluid generates finger
tip-splitting, which is absent in the rectangular geometry for
ordinary fluids.
\end{abstract}

\begin{center}
PACS number(s): 47.20.Gv, 47.20.Ma, 47.20.ky, 03.40.Gc
\end{center}
\end{titlepage}

\section{Introduction}
\label{intro}

The Saffman-Taylor problem~\cite{Saf}, in which two immiscible,
viscous fluids move in a narrow space between the parallel plates of a
Hele-Shaw cell, is a widely studied example of hydrodynamic pattern
formation where interfacial instabilities grow and evolve~\cite{Rev}.
The initially flat interface separating the two fluids is destabilized
by either a pressure gradient advancing the less viscous fluid against
the more viscous one, or by gravity coupling to a density
difference between the fluids.

Ferrofluids, colloidal suspensions of microscopic permanent magnets,
respond paramagnetically to applied fields~\cite{RR}. Because they are
liquids, they flow in response to magnetic forces. Ferrofluids
confined within Hele-Shaw cells exhibit interesting interfacial
instabilities. One of the most beautiful, the labyrinthine
instability, occurs when a magnetic field is applied perpendicular to
the Hele-Shaw cell. Elements of magnetized liquid repel each other,
creating highly branched, intricately fingered structures.

Recent experiments~\cite{Per} examine the Saffman-Taylor instability,
when one of the two fluids is a ferrofluid, in the presence of a
perpendicular magnetic field. The resulting interfacial patterns, in a
rectangular Hele-Shaw cell, are an intriguing superposition of
familiar forms from ordinary viscous fingering and labyrinthine
patterns. Two immediately striking features of the patterns: The
pattern of low viscosity fluid penetrating into high viscosity fluid
is totally unlike the pattern of high viscosity fluid penetrating into
low viscosity fluid; Finger splitting is prevalent, while in general
it is completely absent in zero external magnetic field.

We explain these two phenomena within a perturbative approach known as
mode-coupling theory. Linear stability analysis explains the
instability of an initially flat interface to sinusoidal perturbations
known as modes. In the initial, linear stage of pattern formation,
modes grow or decay independently of each other. One mode, which we
call the ``fundamental'', grows faster than all others. As these
perturbations of the flat interface grow, they evolve through a weakly
non-linear stage, in which modes couple with each other, to the
strongly nonlinear late stages in which a Fourier decomposition of the
interface shape becomes inappropriate.

We carry out our mode-coupling expansion to third order. Linear
stability analysis explains neither interfacial symmetry breaking nor
finger tip-splitting.  At second order, we find the viscosity contrast
$A$ (defined as the difference between the two fluid viscosities
divided by their sum) breaks the symmetry of the interface by
enhancing growth of subharmonic perturbations to the fundamental mode.
This mechanism occurs independently of the applied magnetic field.
At third order we find a mechanism for finger tip-splitting driven by
mutual repulsion of elements of magnetic fluid. In the absence of a
magnetic field, finger tips do not split in rectangular geometry
Hele-Shaw flow.

\section{Hydrodynamics in a Hele-Shaw cell}
\label{hydrodynamics}

This section begins with a discussion of basic hydrodynamic equations
governing the motion of fluids confined within a Hele-Shaw cell,
considering ferrofluid in particular. We present Darcy's law in the
presence of a perpendicular magnetic field, and we discuss boundary
conditions obeyed at the two-fluid interface. Since the basic
equations are well established by previous investigators, we simply
review the chief assumptions and results. This section concludes by
describing our perturbative approach. We introduce a Fourier
decomposition of the interface shape and derive coupled, nonlinear,
ordinary differential equations governing the time evolution of
Fourier amplitudes.

\subsection{Governing equations}
\label{governing}

\begin{figure}
\epsfxsize=300pt \epsfbox{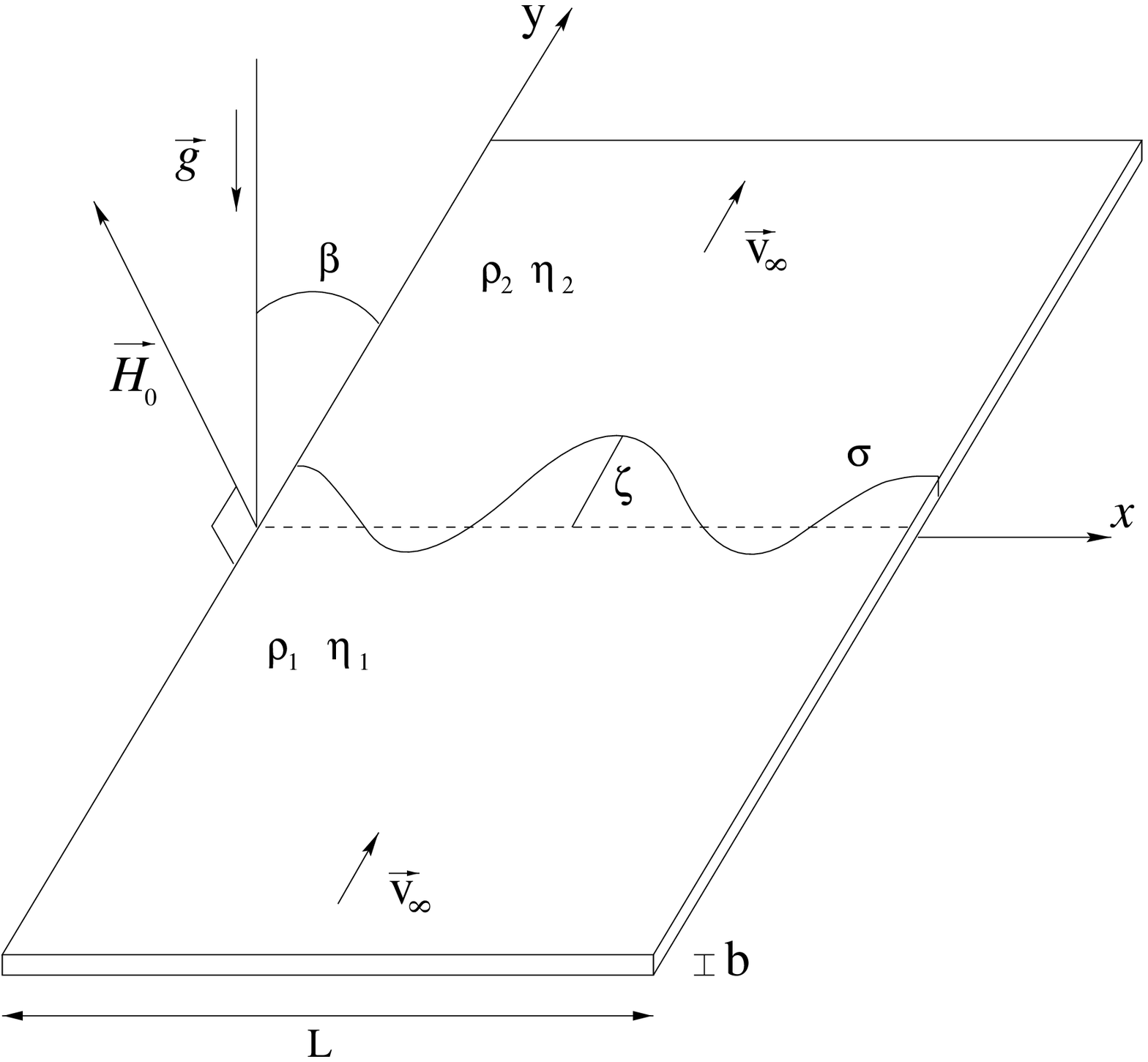}
\caption{Schematic configuration of the rectangular flow geometry. 
The upper fluid is a ferrofluid.}
\label{fig1} 
\end{figure}

Consider two semi-infinite immiscible viscous fluids, flowing in a
narrow gap of thickness $b$, between two parallel plates (see
figure~\ref{fig1}). Denote the densities and viscosities of the lower
and upper fluids, respectively as $\rho_{1}$, $\eta_{1}$ and
$\rho_{2}$, $\eta_{2}$. Between the two fluids there exists a surface
tension $\sigma$. Inject fluid 1 at constant external flow velocity
$\vec v_{\infty}=v_{\infty}\hat{y}$ at $y=-\infty$ and withdraw fluid
2 at the same velocity at $y=+ \infty$.  We describe the system in a
frame moving with velocity $\vec v_{\infty}$, so that the interface
may deform, but it does not displace from $y=0$ (dashed line in
figure~\ref{fig1}) on the average. During the flow, the interface has
a perturbed shape described as $y =\zeta(x,t)$ (solid curve in
figure~\ref{fig1}) over the range $0 \le x \le L$ in the comoving
frame.

In order to include the acceleration of gravity $\vec g$,
we tilt the cell so that the $y$ axis lies at angle $\beta$ from the
vertical direction. To include magnetic forces, we apply a magnetic
field $\vec H_0$ at right angles to the cell. By assumption, the upper
fluid acquires magnetization $\vec M$, while the lower fluid is
nonmagnetic. We consider the limit $L~\rightarrow~\infty$ to simplify
calculations of magnetic forces.

Hydrodynamics of ferrofluids differs from the usual Navier-Stokes
equations through the inclusion of a term representing magnetic force.
Let ${\vec M}$ represent the local magnetization of the ferrofluid,
and note that the force on ${\vec M}$ depends on the gradient of local
magnetic field ${\vec H}$. The local field differs from the applied
field by the demagnetizing field of the polarized ferrofluid.
Restricting our attention to small velocity flows of viscous fluids,
we ignore the inertial terms and write the Navier-Stokes equation for
a single fluid
\begin{equation}
\label{NavierStokes}
- \eta \nabla^{2} \vec u = - \vec \nabla p + 
(\vec M \cdot \vec \nabla) \vec H + \rho {\vec g}
\end{equation}

For the two dimensional geometry of a Hele-Shaw cell, the three
dimensional flow ${\vec u}$, governed by
equation~(\ref{NavierStokes}), may be replaced with an equivalent
two-dimensional flow ${\vec v}$ by averaging over the $z$ direction
perpendicular to the plane of the Hele-Shaw cell. Imposing no-slip
boundary conditions, a parabolic velocity profile and assuming
constant magnetization parallel to $\vec H_0$, one derives Darcy's law
for ferrofluids in a Hele-Shaw cell~\cite{Jac5,Tsebers3},
\begin{equation}
\label{Darcy}
\eta \vec v= -\frac{b^{2}}{12} \left \{ \vec \nabla p - 
\frac{2M}{b} \vec \nabla \varphi -
\rho (\vec g \cdot \hat{y})\hat{y} \right \}.
\end{equation}
The magnetic scalar potential $\varphi$ is evaluated on the top plate
of the cell. The velocity depends on a linear combination of gradients
of $p$ and $\varphi$, so may think of the magnetic scalar potential as
part of an effective pressure. Equation~(\ref{Darcy}) describes
nonmagnetic fluids by simply dropping the terms involving
magnetization.

It is convenient to rewrite equation~(\ref{Darcy}) in terms of
velocity potentials because the velocity field ${\vec v}$ is
irrotational. Since we are interested in perturbations of the velocity
field around a steady flow, we write $\vec v~=~\vec v_{\infty}~-~\vec
\nabla~\phi$, where $\phi$ defines the velocity potential.  Both sides of 
equation~(\ref{Darcy}) are recognized as gradients of scalar fields.
Integrating both sides of equation~(\ref{Darcy}) yields
\begin{equation}
\label{Darcy_phi}
\eta \phi = \frac{b^{2}}{12} \left \{p - \frac{2M}{b} \varphi + 
\rho g y \cos \beta \right \} + \eta v_{\infty} y
\end{equation}
after dropping an arbitrary constant of integration.

Subtract equation~(\ref{Darcy_phi}) for one fluid from the same
equation for the other fluid, then divide by the sum of the two
fluids' viscosities.  This yields an equation for the discontinuity of
velocity potentials valid at the two-fluid interface
\begin{equation}
\label{difference}
A \left ( {{\phi_2 + \phi_1}\over{2}} \right ) + 
\left ( \frac{\phi_2 - \phi_1}{2} \right ) = 
{{b^{2}}\over{12(\eta_{1} + \eta_{2})}} \left (
(p_2 - p_1)
-\frac{2M}{b} \varphi_{s} \right ) + U y.
\end{equation}
The viscosity contrast
\begin{equation}
\label{contrast}
A=\frac{\eta_{2} - \eta_{1}}{\eta_{2} + \eta_{1}}
\end{equation}
will play a key role in interfacial symmetry breaking. $U$ is a characteristic
velocity associated with driving forces,
\begin{equation}
\label{U}
U=\frac{b^{2}(\rho_{2} - \rho_{1}) g \cos \beta}{12(\eta_{1} + \eta_{2})} + 
Av_{\infty}.
\end{equation}

The pressure jump across the interface, $p_1-p_2$ depends on $\kappa$,
the interfacial curvature in the plane of the Hele-Shaw cell. In
general this relationship depends upon the discontinuity of the
viscous stress tensor. Under the assumption of low capillary number flow
we neglect that dependence and write simply
\begin{equation}
\label{delta-p}
p_2-p_1= \sigma \kappa.
\end{equation}

We substitute the pressure jump boundary
condition~(\ref{delta-p}) and also introduce dimensionless variables,
scaling all lengths by the gap size $b$, and all velocities by the
characteristic velocity $\sigma/12(\eta_1+\eta_2)$.
The final equation of motion reads
\begin{equation}
\label{dimensionless}
A \left ( \phi_{2}|_{y=\zeta} + \phi_{1}|_{y=\zeta} \right ) +
\left ( \phi_{2}|_{y=\zeta} - \phi_{1}|_{y=\zeta} \right )
= 
2 \left [U y + 
\kappa - N_{B} {\cal I} \right ]|_{y=\zeta}.
\end{equation}
$N_{B}$ is the dimensionless magnetic Bond number
\begin{equation}
\label{NB}
N_{B}=\frac{2 M^{2} b}{\sigma},
\end{equation}
and the integral
\begin{eqnarray}
\label{Integral}
&{\cal I}= \int_{-\infty}^{\infty} dx' \int_{ \zeta(x')}^{\infty} dy' 
& \left [ \frac{1}{\sqrt{(x - x')^{2} + (y - y')^{2}}} \right. \nonumber\\
&& \left. - \frac{1}{\sqrt{(x - x')^{2} + (y - y')^{2} + 1}} \right ]
\end{eqnarray}
is proportional to the magnetic scalar potential.
Equation~(\ref{dimensionless}) governs the flow for a given interface
shape $\zeta$.

\subsection{Mode-coupling analysis}
\label{mode-coupling}

We begin by representing the net perturbation $\zeta(x, t)$ in the form 
of a Fourier series
\begin{equation}
\label{expansion}
\zeta(x,t)=\sum_{k} \zeta_{k}(t) \exp(ikx),
\end{equation}
where $\zeta_{k}(t)$ denotes the complex Fourier mode amplitudes.
Expansion~(\ref{expansion}) includes a discrete (rather than
continuous) set of modes $k$ because we focus on the interaction of
three particular modes in the subsequent discussion. The $k=0$ mode
vanishes since we are in a comoving frame. The wavevectors are
constrained to lie on the $x$ axis, but can be either positive or
negative.

Now define Fourier expansions for the velocity
potentials $\phi_{i}$, which must obey Laplace's equation 
 $\nabla^{2}\phi_{i}=0$, the boundary conditions 
at $y \rightarrow \pm \infty$, and include the discrete modes $k$
entering the Fourier series~(\ref{expansion}).
The general velocity potentials obeying
these requirements are
\begin{equation}
\label{phi1-2}
\phi_{1}=\sum_{k \neq 0} \phi_{1 k}(t) \exp(|k|y) \exp(ikx),
\end{equation}
and
\begin{equation}
\label{phi2-2}
\phi_{2}=\sum_{k \neq 0} \phi_{2 k}(t) \exp(-|k|y) \exp(ikx).
\end{equation}
In order to substitute expansions~(\ref{phi1-2}) and ~(\ref{phi2-2})
into the equation of motion~(\ref{dimensionless}), we need to evaluate
them at the perturbed interface. For example, expand the lower fluid velocity
potential $\phi_1|_{y=\zeta}$, evaluated at the perturbed interface,
to third order in $\zeta$. Its Fourier transform is
\begin{equation}
\label{Darcy3}
\hat{\phi}_1(k) = \phi_{1k}(t) + 
\sum_{k'} |k'|\phi_{1k'}(t) \zeta_{k - k'} + 
\frac{1}{2} \sum_{k',q} (k')^{2} \phi_{1k'}(t) \zeta_{q}\zeta_{k - k' - q}.
\end{equation}
A similar expression for $\phi_{2}|_{y=\zeta}$ can be easily obtained.
These results define the Fourier transform of the left-hand-side of
equation~(\ref{dimensionless}).

Now we must evaulate the Fourier transform of the right-hand-side of
equation~(\ref{dimensionless}). The curvature in the $x-y$ plane
is~\cite{Dub}
\begin{equation}
\label{curvature}
\kappa=\left ( \frac{\partial^2 \zeta}{\partial x^{2}} \right ) \left [ 1 + \left(\frac{\partial\zeta}{\partial x} \right)^2 \right ]^{-\frac{3}{2}}.
\end{equation}
We expand this up to third order in $\zeta$ and Fourier transform,
\begin{equation}
\label{newequation}
\hat{\kappa}(k) = -k^{2}\zeta_{k} - \frac{3}{2} \sum_{k',q \neq 0} (k')^{2} q [k - k' - q] \zeta_{k'}\zeta_{q}\zeta_{k - k' - q}.
\end{equation}

The expansion to third order in powers of $\zeta$, of the
integral~(\ref{Integral}) related to magnetic scalar potential, is
\begin{eqnarray}
\label{Integral2}
{\cal I}(x) & = & \int_{-\infty}^{\infty} \left [ 
\frac{1}{[(x - x')^{2}]^{1/2}} - \frac{1}{[(x - x')^{2} + 1]^{1/2}} \right ] 
[\zeta(x') - \zeta(x)] dx' \\
            & - & \frac{1}{6} \int_{-\infty}^{\infty} \left [ 
\frac{1}{[(x - x')^{2}]^{3/2}} - \frac{1}{[(x - x')^{2} + 1]^{3/2}} \right ] 
[\zeta(x') - \zeta(x)]^{3} dx'. \nonumber
\end{eqnarray}
When Fourier transformed, the integrals in~(\ref{Integral2}) can be solved 
in terms of
modified Bessel functions~\cite{Gra}
\begin{equation}
\label{BesselRep}
K_{\nu}(k \tau)=\frac{\Gamma(\nu + 1/2)}{k^{\nu} \Gamma(1/2)} (2\tau)^{\nu} \int_{0}^{\infty}\frac{\cos{k x}}{(x^{2} + \tau^{2})^{\nu + 1/2}} dx.
\end{equation}
We define the functions
\begin{equation}
\label{J2}
J(k) \equiv \log \left( \frac{|k|}{2} \right ) + K_{0} (|k|) + C
\end{equation}
with $C$ the Euler constant, and
\begin{eqnarray}
\label{te}
T(k)& \equiv & 3 |k| \Big [ |k| (4 \log 2 - 3 \log 3) + 2 K_{1}(3 |k|) \\
    & - & 4 K_{1}(2 |k|) + 2 K_{1}(|k|) - \frac{2}{3} \Big ], \nonumber
\end{eqnarray}
and write the Fourier transform
\begin{equation}
\label{Iofk}
\hat{{\cal I}}(k)=-2 J(k) \zeta_k +
{{1}\over{6}} \sum_{k',q} T(k-k'-q) \zeta_{k'} \zeta_q \zeta_{k-k'-q}.
\end{equation}
For nonzero $k$, $J(k)$ is positive and $T(k)$ is negative. The
expansion in powers of $\zeta$ can easily be extended to arbitrarily
high order. Tsebers~\cite{Tsebers_comb} presents $\hat{\cal I}(k)$ up
to the fifth order term.

To close equation~(\ref{dimensionless}) we need additional relations
expressing the velocity potentials in terms of the perturbation
amplitudes. To find these, consider the kinematic boundary
condition
relating the interface shape back to the fluid flow. The
condition that the interface move according to the local fluid
velocities is written
\begin{equation}
\label{kinematic}
\frac{\partial \zeta}{\partial t}= 
\left ( \frac{\partial \zeta}{\partial x}\frac{\partial \phi_{i}}{\partial x} 
\right )_{y=\zeta} - 
\left (\frac{\partial \phi_{i}}{\partial y} \right )_{y=\zeta}.
\end{equation}
Expand this to third order in $\zeta$ and
then Fourier transform. Solving for $\phi_{ik}(t)$ consistently to
third order in $\zeta$ yields
\begin{eqnarray}
\label{phi1t-2}
\phi_{1k}(t)& = & -\frac{\dot{\zeta}_{k}}{|k|} + 
\sum_{k'} sgn(kk')\dot{\zeta}_{k'}\zeta_{k - k'} \\
& - & \sum_{k',q} 
\frac{kq}{|k|}sgn(k'q)\dot{\zeta}_{k'}\zeta_{q - k'}\zeta_{k - q} + 
\sum_{k',q}\frac{k'}{|k|} 
\left ( k - q - \frac{k'}{2} \right )
\dot{\zeta}_{k'}\zeta_{q}\zeta_{k - k' - q}\nonumber
\end{eqnarray}
and a similar expression for $\phi_{2k}(t)$. The $sgn$ function equals
$\pm 1$ according to the sign of its argument. The overdot denotes
total time derivative.

Substitute this last expression for $\phi_{1k}(t)$ into
equation~(\ref{Darcy3}) for the Fourier transform of
$\phi_1|_{y=\zeta}$, and again keep only cubic terms in the
perturbation amplitude. Repeat the same procedures for fluid 2. The
velocity potentials have now been eliminated from Darcy's
law~(\ref{dimensionless}), and the differential equation of the
interface is
\begin{eqnarray}
\label{modecoupling}
\dot{\zeta}_{k}& = & \lambda(k)\zeta_{k} + A |k| \sum_{k' \neq 0} \left[ 1 - sgn(kk') \right] \dot{\zeta}_{k'}\zeta_{k - k'} \\
               & + & \sum_{k',q} |k| |q| sgn(k'q) \left[ 1 - sgn(kq) \right] \dot{\zeta}_{k'}\zeta_{q - k'}\zeta_{k - q}\nonumber \\
               & + & \sum_{k',q}k' \left[ k - q - \frac{k'}{2} - \frac{|k'||k|}{2k'} \right ] \dot{\zeta}_{k'}\zeta_{q}\zeta_{k - k' - q}\nonumber \\
               & - & \sum_{k',q} \left [ \frac{1}{6} N_{B} T(k - k' - q) + \frac{3}{2} |k| (k')^{2} q [k - k' - q] \right ] \zeta_{k'}\zeta_{q}\zeta_{k - k' - q}. \nonumber
\end{eqnarray}
Here
\begin{equation}
\label{lambda}
\lambda(k)=|k|[U + 2 N_{B}J(k) - k^{2}]
\end{equation}
is the dimensionless linear growth rate multiplying the first order
term in $\zeta$. The second term in equation~(\ref{modecoupling}) is
second order in $\zeta$, and the remaining terms constitute the third
order contribution.

\section{Weakly nonlinear evolution}
\label{evolution}

This section analyzes the evolution of an interface under the
mode-coupling equation~(\ref{modecoupling}) derived in
section~\ref{mode-coupling}. We systematically examine terms in order
of their strength at the onset of the instability. Thus, we begin by
describing the first order term, which captures the well known linear
instability leading to viscous finger growth. Driving forces causing
the instability include magnetic repulsion within the ferrofluid. We
move on to the second order term, noting the interesting coupling of a
fundamental mode and its own subharmonic. This term is responsible for
finger competition. Magnetic forces do not contribute to this, or any
even order, term. Rather, finger competition depends upon the
viscosity contrast $A$. We conclude our discussion at third order.
Here, we show that finger tips may split due to coupling of a
fundamental mode with its own harmonic. The process depends upon the
presence of magnetic repulsion within the ferrofluid.  It does not
occur without a magnetic field.

\subsection{First order}
\label{first}

\begin{figure} 
\epsfxsize=300pt \epsfbox{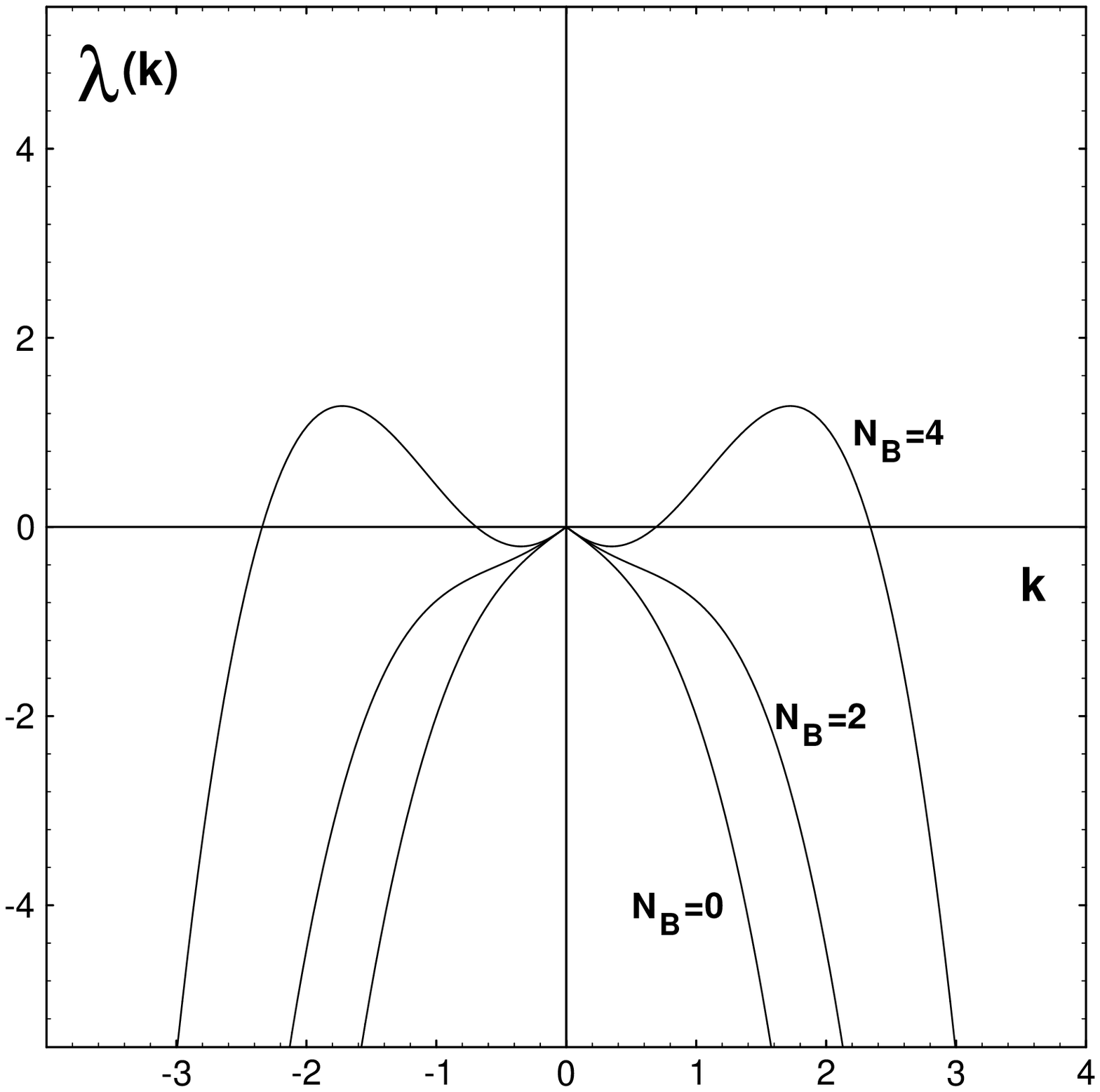} 
\caption{Plot of $\lambda(k)$ for $U=-1$.}
\label{figUm} \end{figure}

\begin{figure}
\epsfxsize=300pt \epsfbox{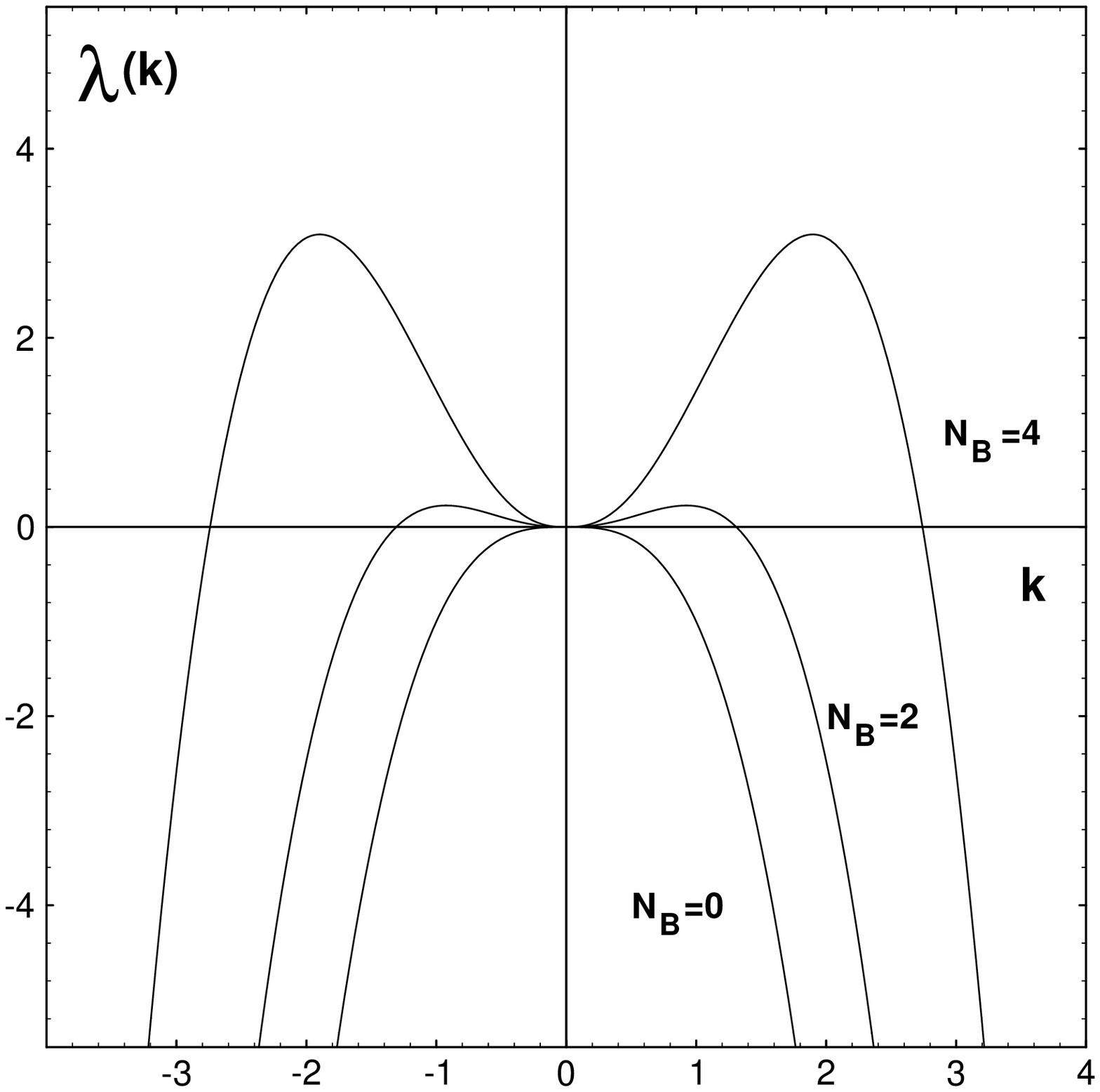} 
\caption{Plot of $\lambda(k)$ for $U=0$.}
\label{figU0} 
\end{figure}

\begin{figure} 
\epsfxsize=300pt \epsfbox{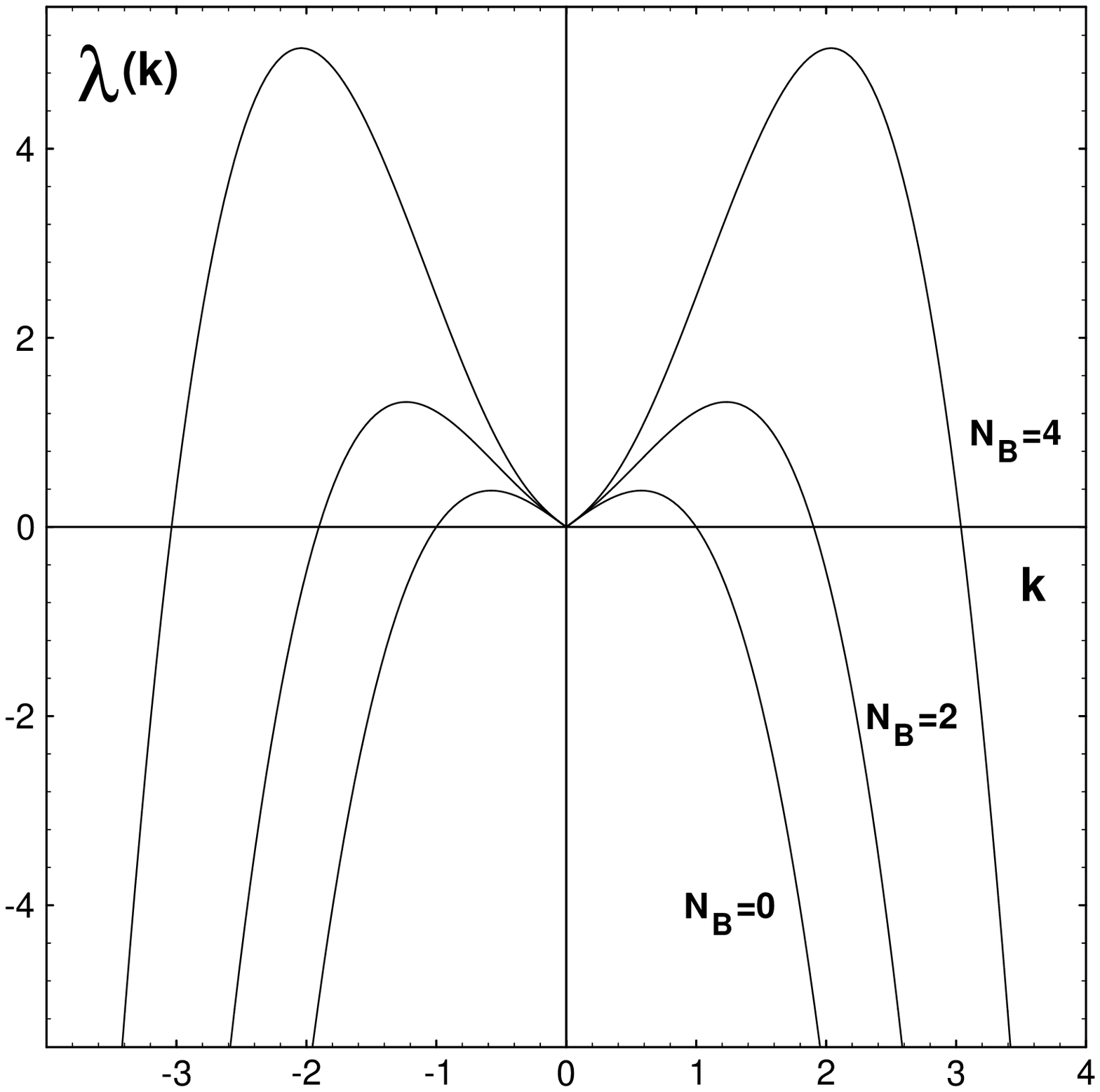} 
\caption{Plot of $\lambda(k)$ for $U=1$.}
\label{figUp} 
\end{figure}

First order in the mode-coupling expansion reproduces conventional
linear stability analysis. Each mode grows or decays independently of
all others, with exponential growth rate $\lambda(k)$ given in
equation~(\ref{lambda}). Positive values of $\lambda(k)$ make a mode
unstable to growth of an initially small perturbation.
Figures~\ref{figUm}~-~\ref{figUp} plot this function for three distinct
cases: $U=-1$, $U=0$ and $U=+1$ respectively. For each value of $U$ we
graph $\lambda(k)$ for three values of the magnetic Bond number, $N_B
= 0,2,4$.

In general these plots display a range of wavenumbers over which
$\lambda(k)>0$. We define two special wavenumbers: $k^{\star}$, the
wavenumber of the fastest growing mode, maximizes $\lambda(k)$; $k_c$,
the threshold wavenumber beyond which all modes are stable, is the
largest wavenumber for which $\lambda(k)$ vanishes. When $U=1$ and
$N_B=0$ we have $k_f=1/\sqrt{3}$ and $k_c=1$. The magnetic field is
destabilizing. As the magnetic Bond number grows, $k^{\star}$ and
$k_c$ shift to the right and modes of higher wavenumber become
unstable. Likewise, for any particular mode $k$, the growth rate
$\lambda(k)$ increases, causing perturbations to grow more rapidly.

To analyze mechanisms of pattern selection, we will focus our
attention on the interaction of one large amplitude perturbation,
which we call the ``fundamental'', with small amplitude perturbations
of its own harmonic and subharmonic. We take the fundamental
wavenumber $k_f=k^{\star}$ of the fastest growing mode. The harmonic
mode $k_h=2k_f$ always lies to the right of the threshold wavenumber
$k_c$, so the harmonic mode is always linearly stable against growth.
The subharmonic mode $k_s=k_f/2$ usually lies in the unstable regime.
If present in the initial conditions it will grow, but less quickly
than the fundamental.

Growth of the fundamental mode creates a sinusoidal oscillation of the
initially flat interface, forming fingers of each fluid penetrating
into the region previously occupied by the other fluid. The interface
is symmetric, with upwards and downwards fingers having identical
length and width. Depending upon the phase of the subharmonic relative
to the fundamental, either the upwards-pointing fingers, or the
downwards pointing fingers may have their length modulated. The
subharmonic can break the up-down symmetry of the growing pattern.
However, within the linear stability analysis, no unique phase of the
subharmonic is favored. Assuming the relative phase is determined by
random perturbation of the flat interface, the growing pattern will
retain {\em statistical} up-down symmetry. For any given initial
condition, symmetry will be broken, but averaged over all initial
conditions, symmetry will remain.

Splitting of fingers is not predicted by linear stability analysis,
because the harmonic mode is required to split fingers, and the
harmonic mode is linearly stable.

\subsection{Second order}
\label{second}

Inspecting the mode coupling equation~(\ref{modecoupling}), we note
that the second order term does not involve magnetic field. We have
previously~\cite{MW_rec} analyzed the role of the second order term in
rectangular flow geometry for non-magnetic fluids. The results are
unchanged, so we will simply recall two essential facts: the second
order term generates finger competition dependent upon the viscosity
contrast $A$; the second order term does not generate finger
tip-splitting. We explain these two points briefly.

Finger competition is linked with the amplitude and phase of the
subharmonic mode. Coupling of the fundamental $k_f$ to the growth of
its subharmonic $k_s$ accelerates growth of the subharmonic, and
selects a preferred phase. The selected phase varies the relative
lengths of fingers of the less viscous fluid penetrating into the more
viscous fluid. Fingers of more viscous fluid penetrating into the less
viscous fluid tend toward equal lengths. The subharmonic therefore
breaks the statistical up-down symmetry of the linear stability
theory.

Finger tip-splitting requires the harmonic mode. In the radial flow
geometry~\cite{MW_rad}, second order terms drive growth of the
harmonic mode $k_h$ despite its linear stability. These terms are
absent due to the rectangular flow geometry. As we explain in the
following section, one must examine the third order terms to
understand growth of the harmonic.

We conclude this discussion with an explanation for the absence of
magnetic field effects at second order. Although the basic equation of
motion~(\ref{Darcy}) is written in terms of forces, it is simplest to
carry out the discussion in terms of energies. Consider the magnetic
energy for a given interface shape $\zeta(x)$. The magnetic energy is
unaffected by rotation of the entire experiment (Hele-Shaw cell and
magnet) by 180$^{\circ}$ around the $x$ axis. Because the ferrofluid
is paramagnetic, the magnetic energy is invariant under reversing the
direction of the applied field. The combination of the two symmetries,
rotation of the experiment followed by reversal of the applied field,
amounts to reversing the sign of the interfacial displacement
$\zeta(x)$. Since the magnetic energy cannot be affected by this
change, it must be an even function of $\zeta(x)$. The magnetic force
is given by the change in magnetic energy with respect to variation in
interfacial shape, so it must be an odd function of $\zeta(x)$.

\subsection{Third order}
\label{third}

This section shows how the magnetic field qualitatively alters the
mechanism for splitting of finger tips. We first review previous
results explaining the general absence of tip splitting in rectangular
geometry flow of ordinary fluids~\cite{MW_rec}. Then we describe a new
mechanism for splitting finger tips in the presence of a magnetic
field.

We consider the influence of the fundamental and sub-harmonic modes on
the growth of the first harmonic. Finger tip-splitting is associated
with the magnitude and phase of the harmonic mode $2k_{f}$. It is
convenient for the subsequent discussions to consider sine and cosine
modes, rather than the complex modes employed in
equation~(\ref{modecoupling}).  Describing the fundamental as a cosine
mode with positive amplitude, we only need to examine the subharmonic
and harmonic cosine modes to analyze finger competition and
tip-splitting. Let $a_k$ denote the amplitude of the cosine mode of
wavenumber $k$.

Earlier papers considered finger tip-splitting for Hele-Shaw flow of
nonmagnetic fluids in the radial~\cite{MW_rad} and
rectangular~\cite{MW_rec} geometries. Of course, the same results hold
for ferrofluids in the absence of applied magnetic fields. The
principal results are as follows. In the radial geometry, a term
proportional to $a_{k_f}^2$ drives growth of the harmonic with the
phase appropriate to split finger tips. In the rectangular geometry,
this second order term is missing. Instead, there is a third order
driving term proportional to $a_{k_f} a_{k_s}^2$. This term is
expected to be too small to split finger tips. There is also a
reduction in the effective stability of the harmonic mode for large
amplitude of the fundamental, but this effect cannot make the harmonic
linearly unstable. Consequently, finger tips do not split under normal
circumstances in the rectangular geometry.

Now we investigate the connection between the applied magnetic field
and the occurrence of finger tip-splitting observed in Hele-Shaw cell
experiments with ferrofluids. For consistency with experimental
results~\cite{Per} we consider the case $U=1$. The equation of motion
for the harmonic mode (neglecting terms of order ${\cal O}
(a_{k_{h}}^{3})$) is
\begin{eqnarray}
\label{akh}
\dot{a}_{k_h} & = & \lambda_{eff} a_{k_h} \\
              & - & 
\left \{ \frac{3}{8} k_{h}k_{s}^2k_{f} \left [ k_{f} + 2k_{s} \right ] 
+ N_{B} k_{h} \left [ \frac{1}{12} T(k_{s}) + \frac{1}{24} T(k_{f}) \right ]  
\right \} a_{k_f}a_{k_s}^{2}. \nonumber
\end{eqnarray}
We incorporate certain third order terms into the effective linear
growth rate
\begin{eqnarray}
\label{lambda_eff}
\lambda_{eff}& = & \lambda(k_h) \\ 
             & + & \left \{ \frac{k_{f}^{2}k_{h}}{2} \left [ \left (k_{f}^{2} + \frac{3}{2}k_{h}^{2} \right ) - 1 \right ] - N_{B} k_{h} \left [ \frac{1}{6} T(k_{f}) + \frac{1}{12} T(k_{h}) \right ]  \right \} a_{k_f}^{2}\nonumber \\ 
             & + & \left \{ \frac{k_{s}^{2}k_{h}}{2} \left [ \left (k_{s}^{2} + \frac{3}{2}k_{h}^{2} \right ) - 1 \right ] - N_{B} k_{h} \left [ \frac{1}{6} T(k_{s}) + \frac{1}{12} T(k_{h}) \right ]  \right \} a_{k_s}^{2}. \nonumber
\end{eqnarray}
In equations~(\ref{akh}) and ~(\ref{lambda_eff}) some terms are explicitly
multiplied by $N_B$ and others are not. We refer to the former as ``magnetic''
terms, and the latter as ``nonmagnetic''. The nonmagnetic terms reproduce the
known mode coupling equation for nonmagnetic fluids~\cite{MW_rec}.

Our mechanism for splitting finger tips focuses on $\lambda_{eff}$.
Initially, this quantity is close to $\lambda(k_h)$, which is strongly
negative because the harmonic mode is stable in the linear theory.
However, $\lambda_{eff}$ is increased by the presence of the modes
$k_f$ and $k_s$, because the coefficients multiplying their squared
amplitudes are positive. To verify this point, recall that $k_f \ge
1/\sqrt{3}$, making the nonmagnetic contribution manifestly positive.
Also, the values of $T(k)$ are negative, making the magnetic
contribution manifestly positive. Since we consider the case in which
$a_{k_f}$ is considerably larger than $a_{k_s}$, the dominant
corrections to the effective growth rate come from the terms
multiplying $a_{k_f}^2$. In the following, we concentrate our
discussion on those terms.

Nonmagnetic terms make the effective growth rate less negative but
cannot make it go positive. The physical reason that these terms do
not make $\lambda_{eff}$ positive can be understood by considering the
contour length of the interface. Introducing the harmonic always
increases the contour length, although the larger the amplitude of the
fundamental, the smaller the increase upon introducing the harmonic.
Multiplying the contour length by the surface tension yields surface
energy that favors minimum contour length. Mathematically, the
nonmagnetic term of order $a_{k_f}^2$ makes $\lambda_{eff}$ less
negative, but if higher orders in perturbation theory were included it
would be evident that $a_{k_f}$ cannot drive $\lambda_{eff}$ positive
without assistance from the magnetic terms.

The terms in $\lambda_{eff}$ that are multiplied by $N_B$ allow
$\lambda_{eff}$ to eventually go positive, permitting growth of the
harmonic. The effective growth rate remains negative up to a threshold
value of $a_{k_{f}}$ for which $\lambda_{eff}=0$. When $a_{k_{f}}$
grows beyond this threshold value, $a_{k_{h}}$ grows rapidly. The
threshold value of $a_{k_f}$ should vary as the inverse square root of
the magnetic Bond number, so tip splitting emerges sooner in strong
magnetic fields.

The harmonic mode enters spontaneously through the third order driving
term proportional to $a_{k_f} a_{k_s}^2$ in equation~(\ref{akh}). As
long as $\lambda_{eff}$ remains negative, this small driving force
should be of little consequence. After $\lambda_{eff}$ goes positive,
however, this term can introduce a harmonic even if none is present in
initial conditions. The existence and phase of the spontaneously
generated harmonic depends on interplay of the fundamental and the
subharmonic.

\begin{figure} \epsfxsize=300pt \epsfbox{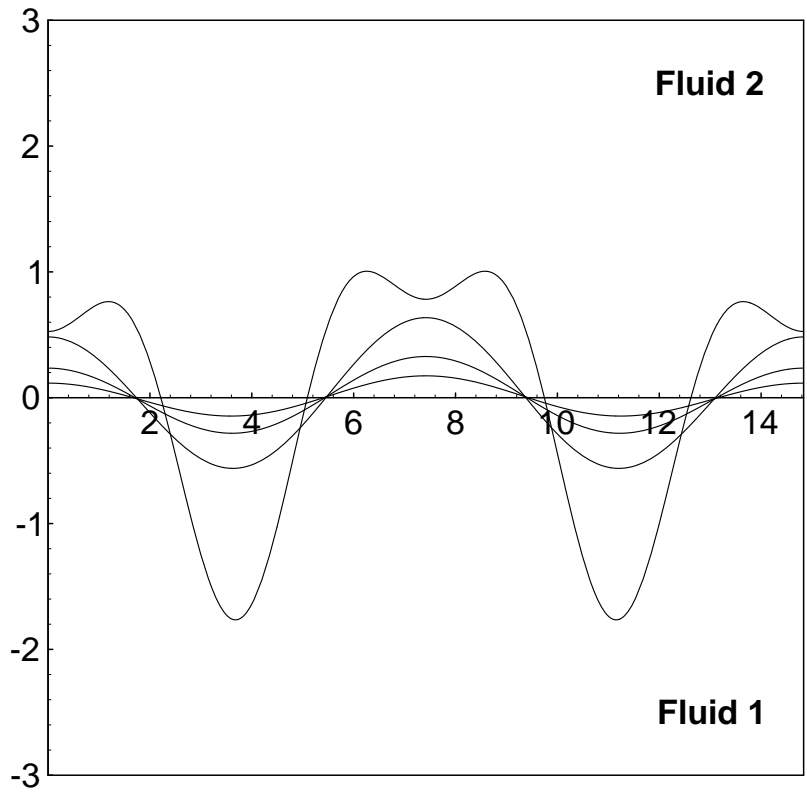} 
\caption{Plot of an evolving interface with an applied magnetic field.}
\label{evol} 
\end{figure}

To illustrate the occurrence of finger tip-splitting when an external
magnetic field is applied, we consider the interaction of modes
$k_{f}$ and $k_{f}/2$ with the forced modes $2k_{f}$ and $3k_{f}/2$.
Mode $3k_{f}/2$ behaves similarly to the sub-harmonic $k_{f}/2$ and
induces more finger competition. In figure~\ref{evol} we plot the
interface evolution using the full solution to third order of
equation~(\ref{modecoupling}). We examine the case in which $U=1$,
assuming that fluid 2 is a ferrofluid. An external magnetic field is
applied ($N_{B}=1.0$) normal to the cell plates. The initial condition
is $a_{k_h}=0.145$ and $a_{k_s}=-1/5~a_{k_f}$. The harmonic
mode is absent initially. Times shown are $t=0,1,2,3$.

The effective harmonic growth rate $\lambda_{eff}$ starts being
strongly negative at $t = 0$. It increases with time, goes through
zero $t \approx 2.16$, and become positive. This process leads to
finger tip-splitting by $t=3$, as shown in figure~\ref{evol}. Finger
tip-splitting only occurs after the fundamental has grown sufficiently
that $\lambda_{eff}$ goes positive. The selected phase of the harmonic
splits the tips of fingers whose length is variable, the less viscous
fingers.

\section{Conclusion}
\label{conclusion}

Several features of the patterns formed in Saffman-Taylor experiments
with ferrofluids can now be explained. First of all, there is the
striking asymmetry of the interface. Since the dense upper fluid is a
glycerine-based ferrofluid with high viscosity, and the less dense
lower fluid is white spirit, we can understand the initial asymmetry
of the interface purely on the basis of the viscosity contrast $A$, as
discussed in section~\ref{second}. Indeed, with the field turned off
the interface is quite asymmetric, with short and wide upwards fingers
of the less viscous fluid and long, thin downward fingers of the more
viscous fluid. Magnetic field effects can exaggerate an already
asymmetric interface, but they cannot break the symmetry by
themselves. It would be of considerable interest to perform the
experiment with immiscible viscosity-matched fluids.

Next, there is the splitting of finger tips which is not normally
observed in rectangular geometry flow. Both upwards and downwards
fingers are split, consistent with a positive value of $\lambda_{eff}$
permitting the growth of harmonics of any phase. The upwards fingers
are more strongly split, however, consistent with the phase preferred
by the driving force in equation~(\ref{akh}). Both are predictions of
our third order analysis in section~\ref{third}. It would be of
interest to examine the relationship between the onset of tip
splitting and the strength of applied magnetic field experimentally.

The above features are explained by our mode coupling theory. Another
notable feature of the patterns is the nearly constant width and
regular spacing of downward pointing fingers of high viscosity fluid.
The constant width is probably the known field-dependent preferred
finger width of the labyrinthine instability~\cite{RR}, proportional
to the plate spacing $b$. Given a directed set of thin fingers,
magnetic forces will drive them towards maximal spacing, resulting in
a regularly spaced array. These issues lie beyond the scope of our low
order mode coupling approach, but may be amenable to more general
forms of weakly nonliner analysis.

\noindent
{\bf Acknowledgments}\\
\noindent
We thank Christiane Caroli for suggesting this research topic, and
the ferrofluid research group at the University of Paris (Pierre et
Marie Curie) for demonstrating the hydrodynamic flows analyzed in this
paper. This work was supported in part by the National Science
Foundation grant No. DMR-9221596. J.A.M. (CNPq reference number
200204/93-9) would like to thank CNPq (Brazilian Research Council) for
financial support. M.W. thanks Leo P. Kadanoff for stimulating
discussions on the subject of viscous fingering in 1983 and later.

\pagebreak

\end{document}